\documentclass[twocolumn,prl,aps,showpacs,superscriptaddress]{revtex4}
\usepackage{xspace,amsmath,amsfonts,amsthm,amssymb,amsbsy,graphicx,color}

\begin{document}

\newtheorem{theo}{Theorem}
\newtheorem{lemma}{Lemma} 

\title{Continuous Measurement of the Energy Eigenstates of a   
Nanomechanical Resonator without a Non-Demolition Probe}

\author{Kurt Jacobs}

\affiliation{Department of Physics, University of Massachusetts at Boston, 100 Morrissey Blvd, Boston, MA 02125, USA}

\affiliation{Quantum Science and Technologies Group, Hearne Institute for Theoretical Physics, Department of Physics and Astronomy, Louisiana State University, 202 Nicholson Hall, Tower Drive, Baton Rouge, LA 70803, USA}

\author{Pavel Lougovski}

\affiliation{Quantum Science and Technologies Group, Hearne Institute for Theoretical Physics, Department of Physics and Astronomy, Louisiana State University, 202 Nicholson Hall, Tower Drive, Baton Rouge, LA 70803, USA}

\author{Miles Blencowe}

\affiliation{Department of Physics and Astronomy, Dartmouth College, Hanover, NH 03755, USA}

\begin{abstract}
We show that it is possible to perform a continuous measurement that continually projects a nano-resonator into its energy eigenstates by employing a linear coupling with a two-state system. This technique makes it possible to perform a measurement that exposes the quantum nature of the resonator by coupling it to a Cooper-pair Box and a superconducting transmission-line resonator. 
\end{abstract}


\maketitle
It is now possible to construct nanomechanical resonators with frequencies on the order of 100 MHz, and quality factors of $10^{5}$~\cite{Cleland98,Craighead00, Zalalutdinov00,Buks01,Huang03, Knobel03,LaHaye04,Badzey05,Naik06}. This 
opens up the exciting prospect of observing quantum behavior in mesoscopic mechanical systems, implementing quantum feedback control in these devices~\cite{Jacobs06xb,Hopkins03}, and exploiting them in technologies for such applications as metrology and information processing~\cite{Cleland04}. The position of these resonators can be monitored by using a single electron transistor (SET) placed nearby~\cite{Blencowe00,Zhang02,Blencowe04}, and such a measurement has recently been realized close to the quantum limit by Schwab {\em et al.}~\cite{LaHaye04,Naik06}. 
However, to observe the quantum nature of a nano-resonator one must measure an  observable that is not linear in the position or momentum of the resonator, and such measurements are considerably more difficult to devise. One approach that has been investigated is to perform a Quantum Non-Demolition (QND) measurement of the resonator energy which would project the resonator into its (discrete) energy eigenstates. This would result in the observation of jumps between these states, a clear signature of quantum behavior. However, such a measurement requires the construction of a non-linear interaction with the resonator, and devising such a coupling with sufficient strength is challenging~\cite{Santamore04,Santamore04b}. 

Here we show that it is possible to perform a measurement that continually projects the resonator onto the basis of energy eigenstates (the Fock basis) using only a linear coupling to a probe system. While the resulting measurement is not a QND measurement, it nevertheless allows a direct observation of the quantum nature of the resonator because it continually projects the system onto the Fock basis. The method exploits the fact that the linear coupling will transfer the effect of a non-linear measurement of the probe onto the resonator, and has similarities with that used in atom optics in the detection of quantum jumps with resonance fluorescence~\cite{Blatt88,Nagourney86,Sauter86,Bergquist86}. The measurement technique we describe is also expected to have applications to state-preparation and feedback control~\cite{Hopkins03}.  

Before we begin we briefly discuss the anatomy of a quantum measurement.  To perform a measurement of an observable $A$ of a quantum system one couples the system to a second ``probe'' system. If one choses an interaction Hamiltonian $H = \hbar \lambda AB$, where $B$ is an observable of the probe, then after a time $t$, this will cause a shift of $\lambda A t$ in the probe observable conjugate to $B$, which we will call $C$. This shift in $C$ can be measured to obtain the value of $A$. The observable $B$ is chosen so that its conjugate observable $C$ can be easily measured directly by an interaction with a classical apparatus. To obtain a continuous measurement of $A$ one proceeds in an analogous fashion,  except that the interaction is kept on, and $C$ is continually monitored. Such a measurement provides a continual stream of information about $A$, and is usually referred to as a {\em continuous measurement}~\cite{JacobsSteck06}. Such a measurement will continually project the system onto the eigenstates of $A$. The measurement is referred to as a QND measurement if $A$ commutes with the system Hamiltonian, so that the system remains in an eigenstate of $A$ once placed there by the measurment~\cite{WMqoptics}. 

We now consider coupling a nano-resonator to a second harmonic oscillator via an interaction linear in the resonators position: $H = \hbar \lambda x_{\mbox{\scriptsize R}} x_{\mbox{\scriptsize P}}$. Here $x_{\mbox{\scriptsize R}} \equiv a + a^\dagger$ is the resonator position, and $x_{\mbox{\scriptsize P}} \equiv b + b^\dagger$ is the position of the probe oscillator which we will take to have the same frequency as the resonator, $\omega_{\mbox{\scriptsize R}}$. This coupling transfers energy between the two oscillators, as well as correlating their phases. If $\lambda \ll \omega_{\mbox{\scriptsize R}}$ then the rotating wave approximation gives $H = \hbar \lambda (ab^\dagger + ba^\dagger)$ which is an explicit interchange of phonons. 

Now consider what happens if we perform a continuous measurement of the energy of the probe oscillator. (This would be a QND measurement of the probe if it were not coupled to the resonator.) Since the energy of the resonator continually feeds into the probe oscillator (and {\em vice versa}), this measurement of the probe must tell us about the energy of the resonator. We should therefore expect the measurement to localize {\em both} the probe and the resonator to their energy eigenstates. This is somewhat surprising from the point of view of the discussion above, since the (linear) position coupling would be expected to transfer phase information to the probe, disturbing the energy eigenstates and projecting instead onto the position basis.  The two arguments can be reconciled by noting that the phase information regarding the system is contained in the phase of the probe, and this phase is continually destroyed by the energy measurement on the probe. Since the probe is projected into an energy eigenstate, the interaction does not imprint a phase back on the resonator from the probe, and there is nothing to prevent it localizing to an energy eigenstate. Nevertheless, the expected disturbance to the Fock states is not completely eliminated, as we shall see, because the linear interaction causes jumps between these states.  

To test the above intuition, we now simulate the evolution of the coupled oscillators. The stochastic master equation (SME) describing their dynamics is
\begin{eqnarray}
  d\rho & = &  -(i/\hbar) [H,\rho]dt  - k[N,[N, \rho ]] dt  \nonumber \\ 
           &    &  + 4k[N\rho + \rho N  - 2\langle N \rangle \rho] (dr - \langle N \rangle dt)  ,
           \label{eq1}
\end{eqnarray}
where $N \equiv b^\dagger b$ is the phonon number operator for the probe, $H = \hbar \lambda (ab^\dagger + ba^\dagger)$, and $k$ is the strength of the energy measurement on the probe. The observers measurement record is $dr=\langle N \rangle dt + dW/\sqrt{8k}$, where $dW$ is Gaussian white noise satisfying the relation $dW^2=dt$~\cite{WienerIntroPaper}. The observer obtains $\rho(t)$ by using her measurement record to integrate Eq.(\ref{eq1}). The simulation is performed using a second order integrator for the deterministic motion, and a simple half-order Newton integrator for the noise term. This involves picking a random Gaussian variable with variance $\Delta t$ at each time-step $\Delta t$. We choose the initial states of the two oscillators as coherent states with mean phonon number $2$, and measure time in units of $k$. Since there is no additional noise appart from that induced by the measurement, we can use the stochastic schr\"{o}dinger equation equivalent to Eq.(\ref{eq1}), which reduces the numerical overhead~\cite{WisemanLinQ}.

\begin{figure}
\leavevmode\includegraphics[width=1.0\hsize]{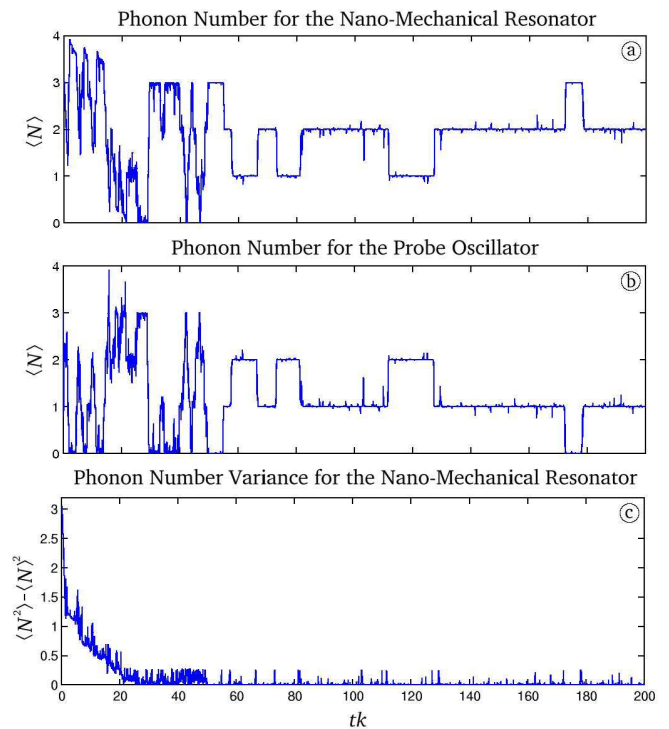}
\caption{Here we plot the evolution of the nanomechanical resonator and the probe oscillator: (a) the energy of the resonator; (b) the energy of the probe; (c) the variance of the energy of the resonator. The interaction strength $\lambda=k/20$ for $tk<50$, and $\lambda=7.5\times 10^{-3}k$ for $tk \geq 50$.} 
\label{fig1}
\end{figure}

The results of the simulation are depicted in Figure~\ref{fig1}. Setting the initial interaction strength to $\lambda=k/20$, we find as expected that the resonator's energy variance reduces essentially to zero at rate $\lambda$. The measurement thus projects the system onto the energy eigenbasis as required. If we start the probe system in a known energy eigenstate (by measuring its energy before we switch on the interaction), then the measurement process also provides full information regarding the initial energy of the resonator, as required of a measurement of energy. However, the interaction causes an additional effect: the two oscillators undergo equal and opposite quantum jumps between their energy eigenstates. (After a time of $t=50/k$ we reduce $\lambda$. This reduces the rate of jumps so that both the jumps and the periods of stability are clearly visible.) This behavior can be understood as follows. The energy measurement tends to keep the resonator and the probe in their energy eigenstates because of the quantum Zeno effect. However, the interaction is continually trying to transfer energy between the two oscillators, and at random intervals this overcomes the quantum Zeno effect and the two oscillators jump simultaneously between energy states.  The jumps are equal and opposite and thus preserve their combined phonon number. The rate of the jumps is determined by the relative size of $\lambda$ and $k$: when $k\gg\lambda$ the jumps are suppressed by the quantum Zeno effect, the energy transfer rate is reduced, and correspondingly the rate of information extraction from the system is also reduced. 

\begin{figure}
\leavevmode\includegraphics[width=0.95\hsize]{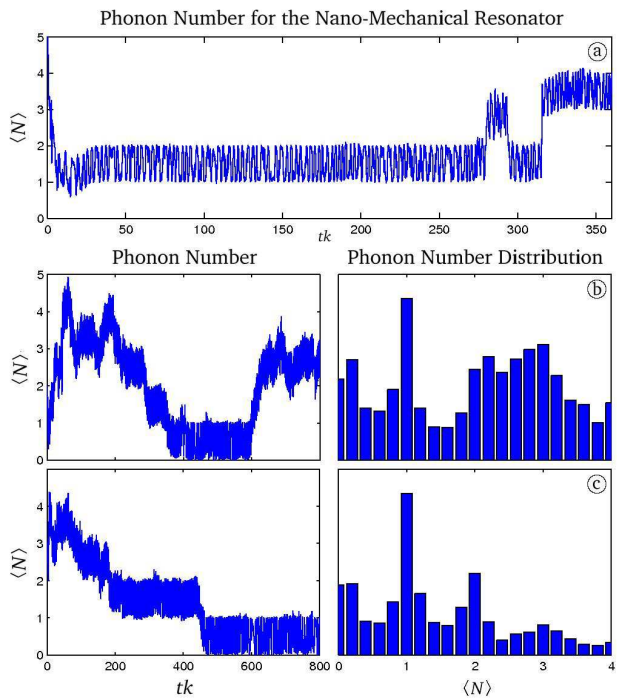}
\caption{Here we plot the evolution of the nanomechanical resonator under a measurement by a Cooper-pair box, as well as histograms of the distribution of the average phonon number over the evolution. (a) Zero temperature ($T=0$) and no damping ($\Gamma=0$); (b) $T = 6 \mbox{mK}$ and $\Gamma = k/500$; (c) $T = 32 \mbox{mK}$ and $\Gamma = k/2500$.}
\label{fig2}
\end{figure}

So far we have been considering a Harmonic oscillator as the probe system. We note now that since a harmonic oscillator truncated to it lowest two energy levels is a two-level system, this suggests that one might be able to use a two-level system as a probe in the same way. This would increase considerably the range of possible experimental realizations. We find that this is indeed the case; a two-level system (TLS) is similarly effective at projecting the nano-resonator onto an energy eigenstate. If we truncate the probe harmonic oscillator to its lowest two levels, then the interaction between the system and probe becomes $H = \lambda  \sigma_x x_{\mbox{\scriptsize R}}$, and the energy measurement on the probe is a measurement of $\sigma_z$. In figure~\ref{fig2}(a), we plot the evolution of the energy of the resonator under such a measurement. Whereas in our previous simulations we assumed that $\lambda \ll \omega_{\mbox{\scriptsize R}} $ and made the rotating wave approximation, here we make no approximation. We choose $k=\omega_{\mbox{\scriptsize R}}/20$, and $\lambda=(3/4)k$, so that there are rapid exchanges of energy between the two systems. Once again the variance of the nano-resonator's energy reduces as required, but this time the resonator jumps (rapidly) between only two adjacent energy levels, since the TLS has only two energy states. We also see a new effect due to the fact that the total number of excitations is no longer preserved by the interaction (because we have not made the rotating wave approximation). Because of this the resonator gets energy kicks from the TLS that are not associated with a flip of the TLS state. These are occasionally sufficient to cause the resonator to jump between phonon states, shifting the offset of the rapid oscillations up or down by one phonon. 

We now show how such an energy measurement can be implemented using a Cooper-Pair-Box (CPB) coupled in turn to a superconducting transmission-line resonator~\cite{Nakamura99}. A CPB is a superconducting island, whose two charge states consist of the presence or absence of a Cooper-pair on this island. If we work at the degeneracy point where the two charge states have the same charging energy, then the free Hamiltonian of the CPB contains only the Josephson tunneling term $E_{\mbox{\scriptsize J}}\sigma_x$. If we place a bias voltage with frequency $\delta$ on the nano-resonator, and place the CPB adjacent to it, we obtain the coupling term $\lambda \cos(\delta t) \sigma_z x_{\mbox{\scriptsize R}}$~\cite{Armour02}. If we then place the CPB in a superconducting resonator (SR), and detune the Josephson tunneling frequency $E_{\mbox{\scriptsize J}}/\hbar$ from the SR frequency $\omega_{\mbox{\scriptsize S}}$ by an amount $\Delta$, then the interaction between the CPB and the superconducting resonator is well approximated by the Hamiltonian $H = \hbar(g^2/\Delta)\sigma_x c^\dagger c$~\cite{Sarovar05}. Here $c$ is the annihilation operator for the SR mode and $g$ is the so-called ``circuit QED'' coupling constant between the CPB and the SR~\cite{Blais04}. This approximation requires that $(E_{\mbox{\scriptsize J}}/\hbar,\omega_{\mbox{\scriptsize S}}) \gg \Delta$ and $\Delta\gg g$, and we set $\delta = E_{\mbox{\scriptsize J}}/\hbar - \omega_{\mbox{\scriptsize R}}$ to bring the resonator-CPB interaction on-resonance. Thus the full Hamiltonian for the nano-resonator, the CPB and the superconducting resonator is 
\begin{equation}
 \frac{\tilde{H}}{\hbar} = \omega_{\mbox{\scriptsize R}}a^\dagger a  + \lambda \cos(\delta t) \sigma_z  x_{\mbox{\scriptsize R}}
 				+ \sigma_x \left[ \frac{E_{\mbox{\scriptsize J}}}{\hbar} + \frac{g^2}{\Delta} c^\dagger c \right] 
				    + \omega_{\mbox{\scriptsize S}}  c^\dagger c .
\end{equation}
This achieves the required configuration in which the nano-resonator is coupled to a CPB via one Pauli operator, and the CPB is coupled to a second probe system via a second Pauli operator. All that is required now is that we use the second probe system (the SR) to perform a continuous measurement of $\sigma_x$.  The interaction term $\hbar(g^2/\Delta)\sigma_x c^\dagger c$ means that the $\sigma_x$ eigenstates of the CPB generate a frequency shift of the SR, which in turn produces a phase shift in the electrical signal carried by a transmission line connected to the SR. Two methods for continuously monitoring this phase shift with high fidelity have been devised by Sarovar {\em et al.}~\cite{Sarovar05}.  If one performs a continuous measurement of the phase of the SR output signal, then one can adiabatically eliminate the SR and obtain an equation describing the continuous measurement of the CPB (this type of adiabatic elimination procedure is detailed in \cite{DJ}). The resulting SME is precisely Eq.(\ref{eq1}), where the Hamiltonian $H$ is replaced by $\tilde{H}$ and the phonon number $N$ is replaced by $\sigma_x$. The important quantity is the final measurement strength $k$ of this $\sigma_x$ measurement. The adiabatic elimination results in the measurement strength $k = (g^4 |\alpha|^2)/(\Delta^2\gamma)$, where $\gamma$ is the decay rate of the SR, and $|\alpha|^2$ is the average number of photons in the SR during the measurement.  The adiabatic elimination requires that $\gamma\gg g^2/\Delta$. We note that this second inequality is merely required to ensure the accuracy of our expression for $k$ --- the measurement can be expected to remain effective without it.  

Readily obtainable values for the circuit QED parameters are $g=\pi \times 10^8~\mbox{s}^{-1}$ and $\gamma$ as low as $6 \times 10^6~\mbox{s}^{-1}$~\cite{Blais04}. A realistic frequency for the nano-resonator is $\omega_{\mbox{\scriptsize R}}/2\pi= 100~\mbox{MHz}$ and for the superconducting resonator is $\omega_{\mbox{\scriptsize S}}/2\pi = 10~\mbox{GHz}$. We choose the CPB frequency so that $\Delta = 4\pi\times 10^{8}~\mbox{s}^{-1}$ and set $g = \Delta/40$, which gives $\omega_{\mbox{\scriptsize S}}/\Delta = 50$. With $\gamma = \pi\times 10^7~\mbox{s}^{-1}$ we then have $\gamma \Delta /g^2$ = 20. With these parameters, choosing even a modest value of $|\alpha|^2 = 2\times 10^3$ provides a measurement strength of $k= 4 \times 10^7 \mbox{s}^{-1}$. The interaction strength $\lambda$ is not a limitation, and can easily be as high as $10^8~\mbox{s}^{-1}$~\cite{Armour02,Hopkins03}.  Nano-resonators typically have quality factors of $Q = 10^5$, giving a damping rate of $\Gamma\approx 10^{4}~\mbox{s}^{-1}$.  

We now turn to the question of observing the quantum nature of the resonator. In this measurement scheme the quantum behavior of the resonator is not manifest in energy jumps resulting from exchanges of excitation number with the CPB; even if the resonator were classical these jumps would occur because the CPB states are discrete. The discrete nature of the resonator states are manifest in the fact that the energy measurement localizes the resonator energy to integer multiples of  $\hbar \omega_{\mbox{\scriptsize R}}$, rather than just any value consistent with the thermal distribution. As a result the rapid oscillations due to excitation exchanges only occur between these discrete values (to within the energy localization induced by the measurement). Further, thermal noise does not cause the oscillator to undergo Brownian motion as it would during a continuous energy measurement on a classical oscillator, but instead induces quantum jumps between the discrete levels. As a result a histogram of $\langle N \rangle$ over time is therefore peaked at integer values, in sharp contrast to the classical case. 

Since we can achieve $k \gg \Gamma$, we would expect to be able to observe the discreteness of the energy levels at low temperatures. We now perform numerical simulations to verify this. These simulations are numerically intensive, so we make the rotating wave approximation, and to include the thermal noise we use an approximation to the Brownian motion master equation~\cite{Caldiera89} that takes the Lindblad form~\cite{Lindblad76}: $\dot{\rho} = \Gamma (\xi+1){\cal D}[a/2]\rho + \Gamma \xi {\cal D}[a^\dagger/2]\rho$. Here $\Gamma$ is the damping rate of the resonator, $\xi=\coth[\hbar\omega_{\mbox{\scriptsize R}}/(2 k_{\mbox{\scriptsize B}}T)]$ (where $T$ is temperature), and ${\cal D}[c]\rho = 2c\rho c^\dagger - c^\dagger c\rho - \rho c^\dagger c$ for any operator $c$.  The CPB is also subject to damping and dephasing, and we include both of these at a rate $\gamma_{\mbox{\scriptsize CPB}} = 1\times 10^{6}~\mbox{s}^{-1}$, which is not far from current values~\cite{Gambetta06}. We plot the results in Figure~\ref{fig2} using the parameters given above with $\lambda = (3/4)k$. Figure~\ref{fig2}(b) shows the results for $\Gamma = k/500$ and $T = 6\;\mbox{mK}$ and Figure~\ref{fig2}(c) for $\Gamma = k/2500$ and $T = 32\;\mbox{mK}$. For each case we plot the histogram of $\langle N \rangle$, and this shows that the peaks at integer values are clearly visible. We also see that the effect of the thermal noise is larger when the resonator is in higher energy eigenstates; as the phonon number increases the peaks are washed out and the behavior of $\langle N \rangle$ becomes indistinguishable from Brownian motion. 

{\em Acknowledgements:} K.J. and P.L. were supported by The Hearne Institute
for Theoretical Physics, The National Security Agency, The Army
Research Office and The Disruptive Technologies Office. M.B. is supported by a NIRT grant from NSF. 


\end{document}